\documentclass[submission, Phys]{SciPost}

\begin{document}

\begin{center}{\Large \textbf{
 Wheeler-DeWitt equation in Null-foliated spacetimes
}}\end{center}

\begin{center}
A. K. Mehta\textsuperscript{1}
\end{center}

\begin{center}
{\bf 1} Indian Institute of Science Education and Research, Pune, India
\\
*abhishek.mehta@students.iiserpune.ac.in
\end{center}

\begin{center}
\today
\end{center}


\section*{Abstract}
{\bf
In this paper, the Wheeler-DeWitt (WDW) equation is derived in null-foliated 4D spacetimes. WDW equation written in null-foliated spacetime presents an enormous simplification compared to the spacelike-foliated spacetime as the null-foliations are 2D. These can be solved exactly and uniquely to give the vertex operators of non-critical strings as a solution. A correspondence is established between null surfaces in 4D to string worldsheet geometry. Attempts are made to derive the physical consequences of this correspondence.
}

\vspace{10pt}
\noindent\rule{\textwidth}{1pt}
\tableofcontents\thispagestyle{fancy}
\noindent\rule{\textwidth}{1pt}
\vspace{10pt}

\section{Introduction and Summary of Results}
\label{sec:intro}
Quantum cosmology is an attempt to apply quantum mechanics to the whole cosmos. The ingredient central to this discussion is the Wheeler-DeWitt equation (WDW), originally derived in 3+1 foliated spacetimes \cite{Wiltshire:1995vk}
\begin{align}
\hat{\mathcal{H}} \Psi=\left[-4 \kappa^{2} \mathcal{G}_{i j k l} \frac{\delta^{2}}{\delta h_{i j} \delta h_{k l}}+\frac{\sqrt{h}}{4 \kappa^{2}}\left(-^{3} R+2 \Lambda+4 \kappa^{2} \hat{T}^{0 \hat{0}}\right)\right] \Psi=0
\end{align}
where $\Psi$ is called the wavefunction of the universe and is a wavefunctional over the space of metric $h_{ab}$ which is the metric induced on the spacelike foliation. $\mathcal{G}$ is the DeWitt supermetric given by
\begin{align}
\mathcal{G}^{(i j)(k l)}=\frac{1}{2} \sqrt{h}\left(h^{i k} h^{j l}+h^{i l} h^{j k}-2 h^{i j} h^{k l}\right)
\end{align}
Solving the above equation is difficult because the metric components and the matter field (here a scalar field) together form the superspace
\begin{align}
\frac{\operatorname{Riem}(\Sigma)}{\operatorname{Diff}_{0}(\Sigma)}\label{supspace0}
\end{align}
where the subscript zero deonotes diffeomorphisms connected to identity and
\begin{align}
\operatorname{Riem}(\Sigma) \equiv\left\{h_{i j}(x), \Phi(x) | x \in \Sigma\right\}
\end{align}
which is infinite dimensional due to the label $x^i$ on the foliation $\Sigma$. The way around this problem is to truncate the degrees of freedom and obtain approximate solutions by considering 'minisuperspaces' where the labels $x^i$'s are dropped and therefore, the superspace becomes 6+1 dimensional. This is justified by arguing that cosmologies are isotropic and homogenous, so finding a solution at a point $x_i \in \Sigma$ is the same as finding a solution at a point $x_j \in \Sigma$, so the position labels can be dropped \cite{Wiltshire:1995vk}. However, exact solutions may be obtained through novel techiniques as done in \cite{Ita:2014kga}. Another problem WDW equation faces is the lack of a probabilistic interpretation of $\Psi$ \cite{Chakraborty:1992pq}. In quantum mechanics (QM), probabilistic interpretation is facilitated by the continutity equation which is not possible here because WDW is of the form $\mathcal{H}\Psi = 0$ while QM is of the form $\mathcal{H}\Psi = -i\partial_t\Psi$ which is another problem faced by WDW i.e. the problem of time \cite{Anderson:2010xm}.\\\\In this paper, attempts have been made to find exact solutions to WDW by exploiting the simplicity of the equation in 2+2 foliated spacetime. In 2+2 foliation the WDW in the case of pure gravity looks like
\begin{align}
-\frac{N_v}{16\pi}\sqrt{\sigma}\bigg[~{}^{(2)}R-2N^{-1}_vD^2N_v+N^{-1}_vD^{A}[N_v(\omega_A-D_A\ln(N_u/N_v))]\bigg]\Psi = 0
\end{align}
which may be treated as the main result of this paper. It can be easily simplified further using the other constraints.
Notice that unlike its 3+1 counterpart the above form has no derivatives in the momenta 
  which is an enormous simplification. The outline of the paper is as follows. Starting from Section \ref{2},  a null foliation of the spacetime is constructed in a manner analogous to the construction of the usual 3+1 foliation. By choosing one of the null directions as time, we provide the Hamiltonian formalism for the null-foliated spacetime, towards the end of the section. In Section \ref{kr}, it was shown that the metric on the null surface was shown to be governed by the Louiville theory. The momentum constraint is shown to be equivalent to the Virasoro constraints and also generates the vertex operators. Consistency with the momentum constraint leads to a quantization of the Louiville theory. And finally, attempts were made to seek out physical consequence of this formalism. Here, the correspondence established betweeen null surfaces and string worldsheet geometry in Section \ref{kr}, was exploited to justify punctured null surfaces i.e. null surfaces with some points excluded, which turns out to be cosmic strings.

\section{Null-foliated spacetimes}\label{2}
An attempt is made to do a 2+2 foliation of our spacetime which can be achieved through \cite{Poisson:2009pwt}
\begin{align}
\sigma_{\mu\nu} = g_{\mu\nu}+2l_{(\mu}n_{\nu)}
\end{align}
where $\sigma$ is the metric on the folia where $l_{\mu}$ and $n_{\nu}$ are null vectors normal to the foliation satisfying 
\begin{align}
l^2 = n^2 = 0 \quad l\cdot n = -1
\end{align}
Analogous to the 3+1 foliation of the spacetime, we define optical scalars $u$ and $v$ such that 
\begin{align}
l_{\mu} = -N_u\partial_{\mu}u \quad n_{\mu} = -N_v\partial_{\mu}v
\end{align}
We would also like to define
\begin{align}
&u^{\mu} = \frac{\partial y^{\mu}}{\partial u} \quad v^{\mu} = \frac{\partial y^{\mu}}{\partial v}\notag\\
&e^{\mu}_{A} = \frac{\partial y^{\mu}}{\partial x^A} 
\end{align}
where $y^{\mu}$ are the spacetime coordinates and $x^A$ are the coordinates on the folia with $A=(1, 2)$. Obviously, $l_{\mu}e^{\mu}_A = n_{\mu}e^{\mu}_A = 0$. Since, $u^{\mu}$ and $v^{\nu}$ must be identified with tangent vectors to null curves. Therefore, we may write the following decomposition
\begin{align}
u^{\mu} = N_un^{\mu}+e^{\mu}_A N_u^{A}\quad v^{\mu} = N_vl^{\mu}+e^{\mu}_A N_v^{A}\label{decomp}
\end{align}
where $N^A_u$ and $N^A_v$ are analogous to the shift. Using the above we may now write the spacetime metric as
\begin{align}
ds^2 = -2N_uN_vdudv+\sigma_{AB}(dx^A+N^A_udu+N^A_vdv)(dx^B+N^B_udu+N^B_vdv)\label{metric}
\end{align}
or more explicitly
\begin{align}
g_{\mu\nu} = 
\begin{pmatrix}
\begin{matrix}N^2_u&-N_uN_v+N_u\cdot N_v\\-N_uN_v+N_u\cdot N_v&N^2_v\end{matrix} &&\begin{matrix}N^A_u\\N^A_v\end{matrix}\\
\begin{matrix}(N^A_u)^T && ~~~~~~~~~~~~~~~~(N^A_v)^T\end{matrix} && \begin{matrix}\sigma_{AB}\end{matrix} 
\end{pmatrix}
\end{align}
where $N^2 \equiv N^iN_i$.\footnote{The vector dual to $N^{A}_u$ is defined as $N^{u}_A$ for notation convinience.} From the above it can be shown that $\sqrt{g} = N_uN_v\sqrt{\sigma}$ (Appendix). For the sake of completeness, the inverse of the metric is also displayed
\begin{align}
g^{\mu\nu} = 
\frac{1}{N_uN_v}\begin{pmatrix}
\begin{matrix}0&&&&&-1\\-1&&&&&0\end{matrix} &&\begin{matrix}-N^A_v\\-N^A_u\end{matrix}\\
\begin{matrix}-(N^A_v)^T && -(N^A_u)^T\end{matrix} && \begin{matrix}N_uN_v\sigma^{AB}+2N^{(A}_uN^{B)}_v\end{matrix} 
\end{pmatrix}\label{im}
\end{align}
With $N_u = N_v$, the metric (\ref{metric}) is in agreement with the form obtained in Eq. (12) of \cite{Brady:1995na}. Similar methods were employed in \cite{Huber:2019yfg} to derive a metric similar to (\ref{metric}). But this paper will stick to the form derived above because of its familiarity with the metric obtained in 3+1 foliation \cite{Poisson:2009pwt}.
\subsection{Gauss-Coddazi relations}
Analogs of the Gauss-Coddazi relations can now be obtained. For a full derivation see the Appendix \ref{A}. Here only the final results are stated
\begin{align}
{}^{(2)}R^{\lambda}_{~~\epsilon\mu\nu} = \sigma^{\lambda}_{\tau}\sigma^{\delta}_{\epsilon}\sigma^{\alpha}_{\mu}\sigma^{\rho}_{\nu}~{}^{(4)}R^{\tau}_{~~\delta\alpha\rho}-K_{\mu}^{~\lambda}\tilde{K}_{\nu\epsilon}-\tilde{K}_{\mu}^{~\lambda}K_{\nu\epsilon}+K_{\nu}^{~\lambda}\tilde{K}_{\mu\epsilon}+\tilde{K}_{\nu}^{~\lambda}K_{\mu\epsilon}
\end{align}
From which we obtain,
\begin{align}
{}^{(2)}R={}^{(4)}R+4~{}^{(4)}R_{\mu\nu}l^{\mu}n^{\nu}-2~{}^{(4)}R_{\mu\nu\rho\sigma}l^{\mu}n^{\nu}l^{\rho}n^{\sigma}-2K\tilde{K}+2K_{\mu\nu}\tilde{K}^{\mu\nu} + \text{t.d.}
\end{align}
where $K_{\mu\nu} = \sigma_{\mu}^{\alpha}\sigma_{\nu}^{\beta}\nabla_{\alpha}l_{\beta}$, $\tilde{K}_{\mu\nu} = \sigma_{\mu}^{\alpha}\sigma_{\nu}^{\beta}\nabla_{\alpha}n_{\beta}$.
Hence, the Einstien Hilbert action thus becomes
\begin{align}
S_{EH} &= \frac{1}{16\pi}\int dudv \bigg\{\int_{\Sigma}d^2xN_uN_v\sqrt{\sigma}\big[{}^{(2)}R+\Lambda-2K\tilde{K}+2K_{\mu\nu}\tilde{K}^{\mu\nu}-2Kn^{\lambda}\nabla_{\lambda}\ln N_v-2\tilde{K}l^{\lambda}\nabla_{\lambda}\ln N_u\notag\\&+\frac{1}{2}|\omega_{A}-D_{A}\ln(N_u/N_v)|^2+2D_{A}(\ln N_u)D^{A}(\ln N_v)\big]\bigg\}\label{EH}
\end{align}
\subsection{Hamiltonian formalism}
We need to find momenta conjugate to $\sigma_{ab}$. So, we define
\begin{align}
\dot{\sigma}_{AB} \equiv \mathcal{L}_{u}\sigma_{AB} = (2N_u \tilde{K}_{AB}+D_AN^{u}_B+D_BN^{u}_A)\label{lie}
\end{align}
The second equality is derived in the Appendix. Hence, we may write
\begin{align}
\frac{\delta \tilde{K}_{AB}}{\delta\dot{\sigma}_{CD}}=\frac{1}{4N_u}(\delta^{C}_A\delta^{D}_B+\delta^{C}_B\delta^{D}_A)
\end{align}
Therefore, the conjugate momenta is now given by
\begin{align}
\pi^{CD} = \frac{\delta \tilde{K}_{AB}}{\delta\dot{\sigma}_{CD}}\frac{\delta S}{\delta \tilde{K}_{AB}} = \frac{1}{16\pi}N_v\sqrt{\sigma}(K^{CD}-K\sigma^{CD}-\sigma^{CD}l^{\lambda}\nabla_{\lambda}\ln N_u)
\end{align}
Therefore, we have 
\begin{align}
\mathcal{H} &= \pi^{CD}\dot{\sigma}_{CD}+\pi^A_{v}\dot{N}^v_A+\pi^v\dot{N}_v-\mathcal{L}\notag\\&= N_u[\pi^vN^A_uD_AN_v-\frac{N_v}{16\pi}\sqrt{\sigma}(~{}^{(2)}R+\Lambda+2D_{A}(\ln N_u)D^{A}(\ln N_v))]+2\pi^{AB}D_AN^u_B\notag\\
&-\frac{8\pi N_uN_v}{\sqrt{\sigma}}\pi^A_v\pi^v_A+\pi^v_A[\partial_vN^A_u+[N_u, N_v]^A]
\end{align}
where t.d. stands for total divergence. For additional details, refer the Appendix \ref{A1}.
\section{Wheeler-De Witt equation}\label{kr}
Implementing the hamiltonian constraint gives
\begin{align}
-\frac{N_v}{16\pi}\sqrt{\sigma}\bigg[~{}^{(2)}R+\Lambda-2N^{-1}_vD^2N_v+N^{-1}_vD^{A}[N_v(\omega_A-D_A\ln(N_u/N_v))]\bigg]= 0\label{WDW}
\end{align}
Notice the enormous simplification in the null-foliated spacetime. Since, there are no time derivatives of $N^A_u, N^A_v$ in (\ref{EH}), therefore, $N^A_u, N^A_v$ are non-dynamical and can be set to anything convinient . Also, there is no momenta conjugate to the 2-metric in the above expression. There are additional constraints which are given by
\begin{align}
&C_A:= \pi^vD_AN_v-2\sigma_{CA}D_B{\pi^{BC}}-\frac{\sqrt{\sigma}}{16\pi}\mathcal{L}_{N_v}(\omega_A-D_A\ln(N_u/N_v)\label{mc}\approx 0\\
&\bar{C}_A:= \frac{\sqrt{\sigma}}{16\pi}\mathcal{L}_{N_u}(\omega_A-D_A\ln(N_u/N_v)\approx 0
\end{align}
where $C_A$ is the momentum constraint. Since, $N^A_u$ can be arbitrarily chosen, the constraint $\bar{C}_A$ implies that
\begin{align}
\omega_A = D_A\ln(N_u/N_v) \label{smplfr}
\end{align}
which simplifies (\ref{WDW}) further to
\begin{align}
\sqrt{\sigma}\big[{}^{(2)}R+\Lambda+J\big] \approx 0\label{WDW2}
\end{align}
where $J \equiv J(N_v)= -2N^{-1}_vD^2 N_v$ and the momentum constraint (\ref{mc}) reduces to
\begin{align}
\pi_vD^AN_v-2D_{B}\pi^{BA} \approx 0\label{mc2}
\end{align}
The uniformization theorem allows one to write the metric on a 2-surface as
\begin{align}
\sigma_{AB} = e^{2\omega}\bar{\sigma}_{AB}
\end{align}
where $\bar{\sigma}$ is a metric of constant curvature, therefore, (\ref{WDW2}) becomes
\begin{align}
{}^{(2)}\bar{R}-2\bar{D}^2\omega+\Lambda e^{2\omega}+\bar{J}\approx 0
\end{align}
This is the equation of motion for \cite{Halyo:2016ofb}, \cite{erbin2015notes}
\begin{align}
S_L[J] = \frac{1}{4\pi}\int d^2x \sqrt{\bar{\sigma}}({}^{(2)}\bar{R}\omega+\bar{D}_A\omega \bar{D}^A\omega+\frac{\Lambda}{2}e^{2\omega}+\bar{J}\omega)\label{LT}
\end{align}
which is the Louiville action with a source $\bar{J}$. 
Consider now the following expression
\begin{align}
D^2N_v = -D_A\tilde{\omega}^A
\end{align} 
where $\tilde{\omega}_B = N_v\omega_{B}$. Since, $\tilde{\omega}_B$ is entirely dependent on $N^A_u, N^A_v$, which can be arbitrarily chosen, therefore, they are chosen such that $D_A\tilde{\omega}^B = 0$, which implies that $D^2N_v = 0$. This choice sets $\bar{J} = 0$ rewriting (\ref{LT}) as
\begin{align}
S_L = \frac{1}{4\pi}\int d^2x \sqrt{\bar{\sigma}}(\omega{}^{(2)}\bar{R}+\bar{D}_A\omega \bar{D}^A\omega+\frac{\Lambda}{2}e^{2\omega})\label{LT2}
\end{align}
The condition $D^2N_v = 0$ can be thought of as an equation of motion for
\begin{align}
S_M = \frac{1}{4\pi\alpha'}\int d^2x \sqrt{\sigma}D_AN_vD^AN_v\label{ST}
\end{align}
One can neatly package $S_L$ and $S_M$ together i.e. if $S_M$ is treated as one-dimensional string theory then $S_L$ is the Jacobian of the path integral measure due to the Weyl transformation \cite{erbin2015notes} \cite{Polchinski:1998rq}
\begin{align}
Z[\sigma] = \int [\mathcal{D}N_v]_{\sigma}[\mathcal{D}b]_{\sigma}[\mathcal{D}c]_{\sigma}e^{-S[\sigma, N_v]-S_{gh}[\sigma, b, c]} = e^{-\frac{c}{6}S_L\big|_{\Lambda=0}}\int [\mathcal{D}N_v]_{\bar{\sigma}}[\mathcal{D}b]_{\bar{\sigma}}[\mathcal{D}c]_{\bar{\sigma}}e^{-S[\bar{\sigma}, N_v]-S_{gh}[\bar{\sigma}, b, c]}\label{PI}
\end{align} 
where $S_{gh}$ is the ghost action and $c = 26-D$ where $D$ is the dimension of the string theory.\footnote{Here, $D=1$.}
\subsection{Virasoro constraints}
The reason attempts are made to interpret $N_v$ as a string mode is because (\ref{mc2}) is equivalent to the Virasoro constraints. (\ref{mc2}) is also the generator of vertex operators in this string theory. In this section, the first claim is demonstrated. 
Consider the smeared form of the momentum constraint 
\begin{align}
\xi[C_A]:=\int ~\xi^{A}C_A=\int \pi_v\xi^BD_BN_v-2\int \xi^BD^A\pi_{AB}=\int \pi_v\mathcal{L}_{\xi}N_v+\int ~ \pi^{AB}\mathcal{L}_{\xi}\sigma_{AB}\label{smear}
\end{align}
Now, one can work in the conformal gauge i.e. $\sigma_{AB} = e^{2\omega}\eta_{AB}$\footnote{Or equivalently $ds^2 = e^{2\omega}dzd\bar{z}$}, so that one can write
\begin{align}
\xi^z = \sum_{n} a_{n} z^{n+1} \quad \xi^{\bar{z}} = \sum_{n} \tilde{a}_n \bar{z}^{n+1} \label{powser}
\end{align}
then
\begin{align}
\xi[C_A] = - \sum_{n}(a_n L_n + \tilde{a}_n \tilde{L}_n)
\end{align}
where
\begin{align}
L_n = \int z^{n+1}(\pi_v\partial_zN_v+\pi^{AB}\partial_z\sigma_{AB}) \quad \tilde{L}_n = \int \bar{z}^{n+1}(\pi_v\partial_{\bar z}N_v+\pi^{AB}\partial_{\bar z}\sigma_{AB})
\end{align}
Now, due to the poisson brackets, (\ref{ham})
\begin{align}
&\{\pi^{AB}(x^A, u, v), \sigma_{CD}(x'^A, u, v')\}_{P.B.} = \delta^A_{(D}\delta^B_{C)}\delta^{(2)}(x-x')\delta(v-v')\\
&\{\pi_v(x^A, u, v), N_v(x'^A, u, v')\}_{P.B.} = \delta^{(2)}(x-x')\delta(v-v')
\end{align}
$(L_n, \tilde{L}_n)$ precisely satisfy the Witt algebra\footnote{Refer to Appendix \ref{C} for calculational details.} and therefore, consistency with the momentum constraint imply
\begin{align}
L_n = \tilde{L}_n = 0
\end{align}
which are precisly the Virasoro constraints. Hence, the momentum constraints is equivalent to the Virasoro constraints. One can, therefore, proceed in a way similar to string theory in order to solve for the Wheeler-DeWitt wave-functional $\Psi$
\begin{align}
(\tilde{L}_n +A\delta_{n, 0})\Psi  = (L_n +A\delta_{n, 0})\Psi = 0 \quad n \geq 0 \label{qviro}
\end{align}
This shows that every null-slice is a CFT.
\subsection{Vertex operator}
Consider the symplectic potential
\begin{align}
\Omega = \int \delta\pi^{AB}\wedge\delta\sigma_{AB} = 2\int\delta\pi\wedge\delta\omega+\int\sqrt{\bar{\sigma}}\delta\bar{\pi}^{AB}\wedge\delta\bar{\sigma}_{AB} 
\end{align}
where $\bar{\pi}^{AB}= e^{2\omega}\pi^{AB}/\sqrt{\bar{\sigma}}$. The uniformization theorem was used to derive the above. Now, by using
\begin{align}
\Gamma^{A}_{BC}= \bar{\Gamma}^{A}_{BC}+\delta^{A}_{B}\bar{D}_C\omega+\delta^{A}_{C}\bar{D}_B\omega-\bar{\sigma}_{BC}\bar{D}^A\omega
\end{align}
it can be shown that
\begin{align}
D_{A}\sqrt{\bar{\sigma}} = -2\sqrt{\bar{\sigma}}\bar{D}_A\omega
\end{align}
Finally, using both of the above identities, it can be shown that
\begin{align}
&D_{A}\pi^{AB} = \bar{\pi}^{AB}D_{A}(e^{-2\omega}\sqrt{\bar{\sigma}})+e^{-2\omega}\sqrt{\bar{\sigma}}(\bar{D}_A\bar{\pi}^{AB}+4\bar{\pi}^{AB}D_A\omega-\frac{\pi}{\sqrt{\bar{\sigma}}}\bar{D}^B\omega)\notag\\&= e^{-2\omega}(\sqrt{\bar{\sigma}}\bar{D}_A\bar{\pi}^{AB}-\pi\bar{D}^B\omega)
\end{align}
Therefore, (\ref{mc2}) can be re-written as 
\begin{align}
\pi_v\bar{D}^BN_v-2(\sqrt{\bar{\sigma}}\bar{D}_A\bar{\pi}^{AB}-\pi\bar{D}^B\omega) \approx 0\label{mc3}
\end{align}
By using the Dirac quantization procedure \cite{Wiltshire:1995vk}
\begin{align}
&\bar{\pi}^{AB} \to \frac{1}{\sqrt{\bar{\sigma}}}\frac{\delta}{\delta\bar{\sigma}_{AB}}\notag\\
&\pi \to \frac{1}{2}\frac{\delta}{\delta\omega}\notag\\
&\pi_v \to \frac{\delta}{\delta N_v}
\end{align}
it is now possible to show that
\begin{align}
V_{\alpha, \beta} = \int d^2x \sqrt{\bar{\sigma}}e^{2\alpha\omega}e^{\beta N_v} \equiv  \int d^2x \sqrt{\bar{\sigma}}\mathcal{O}_{\alpha, \beta}\label{VO}
\end{align}
solves (\ref{mc2}). These look like the dressed vertex operators of 1D string theory. $\alpha, \beta$ are left undetermined because the momentum constraint is a generator of diffeomorphisms while $\alpha, \beta$ is fixed by demnading invariance under Weyl rescalings. Therefore, it will be more correct to say that the momentum constraint is a generator of off-shell vertex operators. However, the quantum version of the Virasoro constraint (\ref{qviro}) demands that (\ref{VO}) be invariant under Weyl rescalings. To see whether the above is invariant under Weyl rescaling, consider the renormalized vertex operator given by\cite{Polchinski:1998rq}\cite{swalg:2008}
\begin{align}
[V]_{\mathrm{r}}=\exp \left(\frac{1}{2} \int d^{2} x d^{2} x^{\prime} \Delta\left(x, x^{\prime}\right) \frac{\delta}{\delta X^{\mu}(x)} \frac{\delta}{\delta X_{\mu}\left(x^{\prime}\right)}\right)V
\end{align}
where
\begin{align}
\Delta\left(x, x^{\prime}\right)=\frac{\alpha^{\prime}}{2} \ln d^{2}\left(x, x^{\prime}\right) \quad X^{\mu} = (N_v, \sqrt{\alpha'}~\omega)
\end{align}
where $d(\sigma, \sigma')$ is the geodesic distance between $\sigma$ and $\sigma'$. Now, the Weyl variation denoted by $\delta_{W}$ of the renormalized operator above is
\begin{align}
\delta_{W}[V]_{\mathrm{r}}=[\delta_{W}V]_{\mathrm{r}}+\frac{1}{2} \int d^{2} x d^{2} x^{\prime} \delta_{W}\Delta\left(x, x^{\prime}\right) \frac{\delta}{\delta X^{\mu}(x)} \frac{\delta}{\delta X_{\mu}\left(x^{\prime}\right)}[V]_{\mathrm{r}}
\end{align}
The Weyl variation corresponding to $\bar{\sigma}_{AB}\to e^{2\lambda}\bar{\sigma}_{AB}$ for the vertex operator derived above is 
\begin{align}
\delta_{W}[V_{\alpha, \beta}]_{\mathrm{r}}=\int d^2x \sqrt{\bar{\sigma}}(2-2\alpha (Q-\alpha)-\alpha'\frac{\beta^2}{2})\delta\lambda[V_{\alpha,\beta}]_{\mathrm{r}}
\end{align}
where \cite{erbin2015notes}, \cite{Polchinski:1998rq}
\begin{align}
\delta_{W}\Delta(x, x) = \alpha'\delta\lambda \quad \delta_{W}\omega = -Q\delta\lambda
\end{align}
was used in the above. Now, for the Weyl variation to vanish
\begin{align}
\alpha(Q-\alpha)+\alpha'\frac{\beta^2}{4} = 1
\end{align}
Consider a representation of the momentum constraint defined by
\begin{align}
\hat{\xi}[C_A]:= e^{-S_{cl}}\xi[C_A]e^{S_{cl}}
\end{align}
where $S_{cl}$ is the action evaluated at the classical configuration. The Dirac quantization procedure, therefore, then implies
\begin{align}
\hat{\xi}[C_A]:= \int ~ T^{AB}\mathcal{L}_{\xi}\sigma_{AB} = \oint dz T_{zz}\xi^{z}+\oint d\bar{z} T_{\bar{z}\bar{z}}\xi^{\bar{z}}
\end{align} 
The functional derivative was taken with respect to the classical field configurations. In the second equality, the conformal gauge was used along with the fact that $T_{z\bar{z}} = 0$ for classical field configurations. By making use of (\ref{powser}), one now obtains a more familiar definition of the Virasoro generators
\begin{align}
L_n = \int dz T_{zz} z^{n+1}  \quad \tilde{L}_n = \int d\bar{z} T_{\bar{z}\bar{z}} \bar{z}^{n+1} 
\end{align}
therefore, further reinforcing the fact that the momentum constraint is equivalent to the Virasoro constraint. One can now use the OPE techniques to obtain the full centrally-charged Virasoro algebra, rederive the vertex operators and restate the quantum version of Virasoro constraint.
\subsection{Consistency with momentum constraint}
In the previous sections, it was demonstrated that the momentum constraint is equivalent to the Virasoro constraints and generate the vertex operators. The momentum constraint is satisfied on every null-slice and, since, the actions $S_L$ and $S_M$ are defined on every null slice, therefore, consistency with the momentum constraint demands that $C_AS_L = C_AS_M = 0$ i.e. the momentum constraint is a generator of symmetry\footnote{Diffeomorphism invariance to be precise.} for these actions and indeed it is so. However, the path integral measure must be invariant under these symmetries, therefore, reconsider (\ref{PI}) as
\begin{align}
Z[\bar{\sigma}] = \int[\mathcal{D}\omega]_{\sigma} e^{-\frac{c}{6}S_L[\bar{\sigma}, \omega]}\int [\mathcal{D}N_v]_{\bar{\sigma}}[\mathcal{D}b]_{\bar{\sigma}}[\mathcal{D}c]_{\bar{\sigma}}e^{-S[\bar{\sigma}, N_v]-S_{gh}[\bar{\sigma}, b, c]}
\end{align}  
where this time the measure for the Louiville field is included as well. To redefine the measure in terms of $\bar{\sigma}$ involves renormalizing the Louiville action i.e. \cite{erbin2015notes}
\begin{align}
Z[\bar{\sigma}] = \int[\mathcal{D}\omega]_{\bar{\sigma}} e^{-\mathcal{S}_L[\bar{\sigma}, \omega]\big|_{\Lambda = 0}}\int [\mathcal{D}N_v]_{\bar{\sigma}}[\mathcal{D}b]_{\bar{\sigma}}[\mathcal{D}c]_{\bar{\sigma}}e^{-S[\bar{\sigma}, N_v]-S_{gh}[\bar{\sigma}, b, c]}
\end{align}  
where $\mathcal{S}_L$ is now given by
\begin{align}
\mathcal{S}_L = \int d^2x \sqrt{\bar{\sigma}}(Q\omega{}^{(2)}\bar{R}+D_A\omega D^A\omega+\frac{\Lambda}{2}e^{2b\omega})\label{LT3}
\end{align}
where 
\begin{align}
Q = b+\frac{1}{b} = \sqrt{\frac{25-D}{6}}
\end{align}
For $D=1$, 
\begin{align}
\mathcal{S}_L = \int d^2x \sqrt{\bar{\sigma}}(2\omega{}^{(2)}\bar{R}+D_A\omega D^A\omega+\frac{\Lambda}{2}e^{2\omega})\label{LT4}
\end{align}
This action can be thought of as the quantum version of $S_L$ i.e. while (\ref{WDW}) describes classical configurations of the 2-metric, the above is the quantized version of the same. This is also needed to define the correlations of the vertex operators derived in the previous section.
\section{Punctured Null-Surfaces}
One can now consider coupling gravity to some external sources. The simplest sources one can couple gravity to are delta-function sources. These delta-function sources can be generated by constructing amplitudes in Louiville theory
\begin{align}
\Psi_{\Omega}[\bar{\sigma}] = \int [\mathcal{D}\omega]~e^{-\mathcal{S}_L}\Omega
\end{align}
where $\Omega$ is a product of $n$ puncture operators\footnote{Puncture operators are primaries of the Louiville theory CFT.} inserted at points $p_i$ on the riemann surface
\begin{align}
\Omega = \prod_{i=1}^n \mathcal{O}_{a_i, 0}(p_i) 
\end{align}
These puncture operators contribute the following source term to the action \cite{zamolodchikov2007lectures}
\begin{align}
J = \frac{1}{\sqrt{\bar{\sigma}}}\sum^n_{i=1}a_i\delta^{(2)}(z-p_i)
\end{align}
In presence of this source, the equation of motion is given by
\begin{align}
2\bigg[{}^{(2)}\bar{R}+\frac{1}{\sqrt{\bar{\sigma}}}\sum^n_{i=1}\frac{a_i}{2}\delta^{(2)}(z-p_i)\bigg]-2\bar{\nabla}^2\omega+\Lambda e^{2\omega}=0
\end{align}
The above represents the equation for a riemann metric with $n$-punctures \cite{zamolodchikov2007lectures}. The simplest solution is obtained when $\Lambda = 0$ and $\bar{\sigma}_{AB} = \eta_{AB}$ , which reduces the above to
\begin{align}
\sum^n_{i=1}a_i\delta^{(2)}(z-p_i)-2\bar{\partial}\partial\omega=0
\end{align}
This has a simple solution
\begin{align}
\omega = \frac{1}{\pi}\sum_{i = 1}^na_i\ln|z-p_i|
\end{align}
which means the metric on the 2-surface is 
\begin{align}
\sigma_{AB} = \exp( \frac{1}{\pi}\sum_{i = 1}^na_i\ln|z-p_i| )\delta_{AB}
\end{align}
The equation of motion for $N_v$ becomes $\partial\bar{\partial} N_v = 0$. Therefore, $N_u = N_v =1$ and $N^A_v = N^A_u = 0$ is the simplest choice which satisfies (\ref{smplfr}). Therefore, the complete 4-metric looks like
\begin{align}
ds^2 = -2dudv+\exp( \frac{1}{\pi}\sum_{i = 1}^na_i\ln|z-p_i| )dzd\bar{z}\label{cst}
\end{align}
This is a flat metric whose null hypersurfaces have $n$ punctures on it. The above metric is precisely the metric for multiple cosmic strings \cite{Aliev:1998ae}. $a_i$'s which are the momenta of the Louiville modes, now have the interpretation of mass per unit length of the string. 
\section{Discussion}
In the past few sections, an attempt was made to find exact solutions to the Wheeler De-Witt equations by moving to 2+2 foliations. By doing so, it was possible to write a WDW that was independent of the conjugate momenta. After some simplification using the other constraints, it was understood that the WDW corresponds to the equation of motion of the classical Louiville theory. It was also shown that the momentum constraint is equivalent to the Virasoro constraints and was a generator of vertex operators. Later, the classical Louiville theory was quantized so that the correlations of the primaries in the theory can be defined. The fact that the momentum constraint is equivalent to the Virasoro constraint points to a correspndence between null surfaces and string worldsheet geometry. This correspondence manifested itself with the metric on the null-surface being the metric of a string worldsheet and one of the degrees of freedom of the full 4-metric being the string mode. In the end, correlations of primaries in the Louiville theory was used to induce delta-function source terms in the theory. In presence of these delta-function sources, the equation of motion describes a punctured riemann surface. This punctured riemann surface in the context of null foliations give rise to punctured null surfaces. The cosmic string metric is the simplest metric where one encounters punctured null surfaces. It will be interesting to look at punctured null surfaces of other topologies and undertsand what they mean physically. Other obvious future directions one may now take are finding a similar correspondence with superstring theory and finding other physical consequences of this correspondence.
\section*{Acknowledgement}
I would like to thank Suneeta Varadarajan, Sunil Mukhi, Manu Paranjape, Charles G Torre, Steven Carlip and Hermann Nicolai for fruitful discussions. I also would like to thank Prof. Sachin Jain without whose supervision this project would not have been possible. I also acknowledge the support of CSIR-UGC (JRF) fellowship (09/936(0212)/2019-EMR-I). Finally, I would like to acknowledge our debt to the steady support of the people of India for research in the basic sciences. 

\begin{appendix}
\section{Derivation of Gauss-Coddazi relations}\label{A}
Consider a null hypersurface with two null normals $l^{\mu}$ and $n^{\mu}$, which implies that the metric on the hypersurface is
\begin{align}
\sigma_{\mu\nu} = g_{\mu\nu}-2l_{(\mu}n_{\nu)}\label{pro}
\end{align}
which is also taken to be the orthogonal projector. Given a $\mathcal{V}^{\mu} \in \mathcal{T}(\mathcal{M})$, we have $(\sigma\cdot \mathcal{V})_{\nu} = \mathcal{V}_{\nu}+n_{\nu}(\mathcal{V}\cdot l)+l_{\mu}(\mathcal{V}\cdot n)\in \mathcal{T}(\Sigma)$ where $\Sigma$ is the null slice, therefore, $\sigma: \mathcal{T}(\mathcal{M})\to\mathcal{T}(\Sigma)$. Hence, any $p$-form belonging to $\mathcal{T}(\mathcal{M})^p$ maybe projected to $\mathcal{T}(\Sigma)^p$ \cite{Wald:1984rg}. With this knowledge, consider a vector $V^{\mu} \in \mathcal{T}(\Sigma)$ . Then we may write
\begin{align}
D_{\mu}D_{\nu}V^{\lambda} = \sigma_{\mu}^{\alpha}\sigma_{\nu}^{\beta}\sigma_{\sigma}^{\lambda}\nabla_{\alpha}(D_{\beta}V^{\sigma}) = \sigma_{\mu}^{\alpha}\sigma_{\nu}^{\beta}\sigma_{\sigma}^{\lambda}\nabla_{\alpha}(\sigma_{\beta}^{\rho}\sigma^{\sigma}_{\tau}\nabla_{\rho}V^{\tau})
\end{align}
where $D$ is the covariant derivative with respect to $\sigma_{\mu\nu}$. Now, by making use of (\ref{pro}), we obtain
\begin{align}
D_{\mu}D_{\nu}V^{\lambda} = (\tilde{K}_{\mu\nu}l^{\rho}+K_{\mu\nu}n^{\rho}\sigma^{\lambda}_{\tau})\nabla_{\rho}V^{\tau}-\sigma_{\nu}^{\rho}(K_{\mu}^{\lambda}\nabla_{\rho}n_{\tau}+\tilde{K}_{\mu}^{\lambda}\nabla_{\rho}l_{\tau})V^{\tau}+\sigma_{\mu}^{\alpha}\sigma_{\nu}^{\rho}\sigma_{\tau}^{\lambda}\nabla_{\alpha}\nabla_{\rho}V^{\tau}
\end{align}
where $K_{\mu\nu} = \sigma_{\mu}^{\alpha}\sigma_{\nu}^{\beta}\nabla_{\alpha}l_{\beta}$, $\tilde{K}_{\mu\nu} = \sigma_{\mu}^{\alpha}\sigma_{\nu}^{\beta}\nabla_{\alpha}n_{\beta}$ are the Weingarten maps respect to the null normals on the hypersurface and are, hence, self-adjoint.
Therefore, antisymmetrizing in $\mu$, $\nu$ gives
\begin{align}
{}^{(2)}R^{\lambda}_{~~\epsilon\mu\nu} = \sigma^{\lambda}_{\tau}\sigma^{\delta}_{\epsilon}\sigma^{\alpha}_{\mu}\sigma^{\rho}_{\nu}~{}^{(4)}R^{\tau}_{~~\delta\alpha\rho}-K_{\mu}^{~\lambda}\tilde{K}_{\nu\epsilon}-\tilde{K}_{\mu}^{~\lambda}K_{\nu\epsilon}+K_{\nu}^{~\lambda}\tilde{K}_{\mu\epsilon}+\tilde{K}_{\nu}^{~\lambda}K_{\mu\epsilon}
\end{align}
Taking traces to obtain
\begin{align}
{}^{(2)}R &= \sigma^{\rho\delta}\sigma^{\alpha}_{\tau}~{}^{(4)}R^{\tau}_{~~\delta\alpha\rho}-2K\tilde{K}+2K_{\mu\nu}\tilde{K}^{\mu\nu}\notag\\&={}^{(4)}R+4~{}^{(4)}R_{\mu\nu}l^{\mu}n^{\nu}-2~{}^{(4)}R_{\mu\nu\rho\sigma}l^{\mu}n^{\nu}l^{\rho}n^{\sigma}-2K\tilde{K}+2K_{\mu\nu}\tilde{K}^{\mu\nu} 
\end{align}
Since,
\begin{align}
{}^{(4)}R_{\mu\nu}l^{\mu}n^{\nu} &= \frac{1}{2}(l^{\nu}(\nabla_{\mu}\nabla_{\nu}-\nabla_{\nu}\nabla_{\mu})n^{\mu}+(l\leftrightarrow n))\notag\\
&=\frac{1}{2}(\nabla_{\mu}(l^{\nu}\nabla_{\nu}n^{\mu}-l^{\mu}\nabla_{\nu}n^{\nu})+(l\leftrightarrow n))+\nabla_{\mu}l^{\mu}\nabla_{\nu}n^{\nu}-\nabla_{\mu}l^{\nu}\nabla_{\nu}n^{\mu}\notag\\
&=\text{t.d.}+K\tilde{K}-K_{AB}\tilde{K}^{AB}-Kl^{\beta}n^{\alpha}\nabla_{\alpha}n_{\beta}-\tilde{K}n^{\beta}l^{\alpha}\nabla_{\alpha}l_{\beta}+\omega^{\rho}l^{\mu}\nabla_{\rho}n_{\mu}-2(n^{\sigma}\nabla_{\sigma}n_{\mu})(l^{\rho}\nabla_{\rho}l^{\mu})\label{td}
\end{align}
\begin{align}
{}^{(4)}R_{\mu\nu\rho\sigma}l^{\mu}n^{\nu}l^{\rho}n^{\sigma} &= (l^{\mu}l^{\rho}n^{\sigma}(\nabla_{\mu}\nabla_{\nu}-\nabla_{\nu}\nabla_{\mu})n^{\mu}+(l\leftrightarrow n))\notag\\
&=\nabla_{\rho}(l^{\mu}l^{\rho}n^{\sigma}\nabla_{\sigma}n_{\mu})+\nabla_{\rho}(n^{\rho}n^{\mu}l^{\sigma}\nabla_{\sigma}l_{\mu})-Kl^{\mu}(n^{\sigma}\nabla_{\sigma}n_{\mu})-\bar{K}n^{\mu}(l^{\sigma}\nabla_{\sigma}l_{\mu})\notag\\
&+\omega^{\rho}l^{\mu}\nabla_{\rho}n_{\mu}+(n^{\sigma}\nabla_{\sigma}l_{\mu})(l^{\rho}\nabla_{\rho}n^{\mu})-3(n^{\sigma}\nabla_{\sigma}n_{\mu})(l^{\rho}\nabla_{\rho}l^{\mu})\label{td1}
\end{align}
The following identities are stated without derivation
\begin{align}
& n^{\mu}\nabla_{\mu}n_{\nu} = n_{\nu} n^{\mu}\nabla_{\mu}\ln N_v\label{id1}\\
&l^{\mu}\nabla_{\mu}l_{\nu} = l_{\nu} l^{\mu}\nabla_{\mu}\ln N_u\label{id2}\\
&n^{\sigma}\nabla_{\sigma}l_{\mu} = \frac{1}{2}(\omega_{\mu}+D_{\mu}\ln(N_u N_v)-n_{\mu}l^{\sigma}\nabla_{\sigma}\ln N_u-l_{\mu}n^{\sigma}\nabla_{\sigma}\ln N_v)\label{id3}\\
&n^{\sigma}\nabla_{\sigma}l_{\mu} = \frac{1}{2}(-\omega_{\mu}+D_{\mu}\ln(N_u N_v)-n_{\mu}l^{\sigma}\nabla_{\sigma}\ln N_u-l_{\mu}n^{\sigma}\nabla_{\sigma}\ln N_v)\label{id4}\\
&\omega^{\mu}n^{\sigma}\nabla_{\mu}l_{\sigma} = \frac{1}{2}(\omega^{A}\omega_{A}-\omega^{A}D_{A}\ln(N_u/N_v))-2(n^{\sigma}\nabla_{\sigma}n_{\mu})(l^{\rho}\nabla_{\rho}l^{\mu})\label{id5}
\end{align}
where $\omega = [n, l]$ and where t.d. stands for the total divergences. This leads to
\begin{align}
{}^{(2)}R &={}^{(4)}R+2K\tilde{K}-2K_{\mu\nu}\tilde{K}^{\mu\nu}+2Kn^{\lambda}\nabla_{\lambda}\ln N_v+2\tilde{K}l^{\lambda}\nabla_{\lambda}\ln N_u\notag\\&-\frac{1}{2}\omega^{A}\omega_{A}+\omega^{A}D_{A}\ln(N_u/N_v)-\frac{1}{2}|D_{A}\ln (N_uN_v)|^2 +\text{t.d.}\label{id6}
\end{align}
Similar Gauss-Coddazi relations for null foliated spacetimes were also derived in \cite{Brady:1995na} using similar techniques.
\subsection{Hamiltonian formulation}\label{A1}
To describe a Hamiltonian, we require something akin to a time derivative. The $u$ direction is taken as 'time'. Hence, the defintion
\begin{align}
\dot{\sigma}_{AB} \equiv \mathcal{L}_{u}\sigma_{AB} = e^{\mu}_Ae^{\nu}_B\mathcal{L}_{u}g_{\mu\nu}
\end{align}
since, $\mathcal{L}_{u}e^{\mu}_A = 0$. Hence, we have
\begin{align}
\dot{\sigma}_{AB}  = e^{\mu}_Ae^{\nu}_B(\nabla_{\mu}u_{\nu}+\nabla_{\nu}u_{\mu}) = (2N_u \tilde{K}_{AB}+N^{u}_{B|A}+N^{u}_{A|B})\label{mom}
\end{align}
where in the second equality (\ref{decomp}) was used. The lagrangian density is given by
\begin{align}
&\mathcal{L} =\frac{1}{16\pi}\sqrt{g}{}^{(4)}R= \frac{1}{16\pi}N_uN_v\sqrt{\sigma}\big[{}^{(2)}R-2K\tilde{K}+2K_{\mu\nu}\tilde{K}^{\mu\nu}-2Kn^{\lambda}\nabla_{\lambda}\ln N_v-2\tilde{K}l^{\lambda}\nabla_{\lambda}\ln N_u\notag\\&+\frac{1}{2}|\omega_{A}-D_{A}\ln(N_u/N_v)|^2+2D_{A}(\ln N_u)D^{A}(\ln N_v)\big]
\end{align}
To compute the hamiltonian density, another identity is needed
\begin{align}
\omega_{A} = \frac{1}{N_uN_v}(N^{B}_uD_B N^v_A-N^{B}_vD_B N^u_A)\equiv -\frac{1}{N_uN_v}\mathcal{L}_{N_v}N^A_u\label{id7}
\end{align}
where $-\mathcal{L}_{N_v}N^A_u = [N_u, N_v]_A = N^B_uD_BN^v_A-N^B_vD_BN^u_A$. Now, we can compute the following conjugate momenta
\begin{align}
&\pi^{CD} = \frac{1}{16\pi}N_v\sqrt{\sigma}(K^{CD}-K\sigma^{CD}-\sigma^{CD}l^{\lambda}\nabla_{\lambda}\ln N_u)\label{PC}\\
&\pi^v = -\frac{\sqrt{\sigma}}{8\pi}K\\
&\pi^u = \pi^u_A = \pi^v_A = 0\label{pc}
\end{align}
(\ref{pc}) are the primary constraints. Therefore, the hamiltonian density becomes
\begin{align}
&\mathcal{H} = \pi^{CD}\dot{\sigma}_{CD}+\pi^v\dot{N}_v-\mathcal{L} \notag\\&= \pi^vN^A_uD_AN_v-[\frac{N_uN_v}{16\pi}\sqrt{\sigma}(~{}^{(2)}R+2D_{A}(\ln N_u)D^{A}(\ln N_v)+\frac{1}{2}|\omega_{A}-D_{A}\ln(N_u/N_v)|^2)]+2\pi^{AB}D_AN^u_B\label{ham}
\end{align}
Now, computing the secondary constraints give
\begin{align}
&C:=\dot{\pi}^u = \{\pi^u, \mathcal{H}\}_{P.B.} = -\frac{N_v}{16\pi}\sqrt{\sigma}\bigg[~{}^{(2)}R-2N^{-1}_vD^2N_v+N^{-1}_vD^{A}[N_v(\omega_A-D_A\ln(N_u/N_v))]\bigg]\approx 0\\
&C_A:=\dot{\pi}^u_A = \{\pi^u_A, \mathcal{H}\}_{P.B.} = \pi^vD_AN_v-2\sigma_{CA}D_B{\pi^{BC}}-\frac{\sqrt{\sigma}}{16\pi}\mathcal{L}_{N_v}(\omega_A-D_A\ln(N_u/N_v)\approx 0\\
&\bar{C}_A:=\dot{\pi}^{v}_A =  \{\pi^v_A, \mathcal{H}\}_{P.B.} = \frac{\sqrt{\sigma}}{16\pi}\mathcal{L}_{N_u}(\omega_A-D_A\ln(N_u/N_v)\approx 0\label{sc2}
\end{align}
These are the hamiltonian and momentum constraints, respectively analogous to the 3+1 case \cite{Giulini:2015qha}.
\subsection{In presence of matter}\label{A2}
In the presence of scalar matter, the secondary constraints undergo slight modifications due to (\ref{H2}). They are as follows
\begin{align}
&C:=\dot{\pi}^u = \{\pi^u, \mathcal{H}\}_{P.B.} = -\frac{N_v}{16\pi}\sqrt{\sigma}(~{}^{(2)}R-\frac{1}{2}D_A\tilde{\Phi}D^A\tilde{\Phi}-\mathcal{V}(\tilde{\Phi})-2N^{-1}_vD^2N_v)\approx 0\\
&C_A:=\dot{\pi}^u_A = \{\pi^u_A, \mathcal{H}\}_{P.B.} = \pi^vD_AN_v-(2\sigma_{CA}D_B{\pi^{BC}}-D_{A}\Phi\pi_{\Phi})-\frac{\sqrt{\sigma}}{16\pi}\mathcal{L}_{N_v}(\omega_A-D_A\ln(N_u/N_v)\approx 0\label{SC2}\\
&\bar{C}_A:=\dot{\pi}^{v}_A =  \{\pi^v_A, \mathcal{H}\}_{P.B.} = \frac{\sqrt{\sigma}}{16\pi}\mathcal{L}_{N_u}(\omega_A-D_A\ln(N_u/N_v)\approx 0\label{SC3}
\end{align}
\section{Going from 3+1 foliation to 2+2 foliation}\label{B}
Consider a 3+1 foliated spacetime manifold $\mathcal{M}$ with a metric $g$. The spacelike-foliation $\Sigma$ will have a 2D timelike boundary $\mathcal{B}$. The induced metric $h$ on $\Sigma$ folitation obey 
\begin{align}
h_{\mu\nu}= m_{\mu}m_{\nu}+g_{\mu\nu}
\end{align} 
where $m$ is the timelike normal of $\Sigma$ normalized such that $m^2 = -1$\cite{Poisson:2009pwt}. Now, coming back to the timelike boundary of the foliation, the induced metric $\sigma$ on $\mathcal{B}$
\begin{align}
\sigma_{\mu\nu} = m_{\mu}m_{\nu}-r_{\mu}r_{\nu}+g_{\mu\nu}
\end{align} 
where $r$ is the spacelike normal of $\mathcal{B}$ and is also tangent to $\Sigma$ such that $r^2 = 1, r\cdot m = 0$\cite{Poisson:2009pwt}. Now, given a timelike vector $m$ and a spacelike vector $r$. One can construct
\begin{align}
l^{\mu} = \frac{r^{\mu}+m^{\mu}}{\sqrt{2}} \quad n^{\mu} = \frac{m^{\mu}-r^{\mu}}{\sqrt{2}}\label{null}
\end{align}
such that $l^2=n^2=0$ and $l\cdot n = -1$. Using the above construction, we can rewrite the induced metric on $\mathcal{B}$ as
\begin{align}
\sigma_{\mu\nu} = l_{\mu}l_{\mu}+n_{\mu}n_{\nu}+g_{\mu\nu}
\end{align}
which is precisely the induced metric on a null-foliation of a null-foliated spacetime. Therefore, given a 3+1 foliation, it is possible to construct a 2+2 foliation this way.
\section{Constraint algebra}\label{C}
\begin{align}
-2\int\xi_BD_A\pi^{AB}=\int ~ \pi^{AB}\mathcal{L}_{\xi}\sigma_{AB}
\end{align}
is demonstrated below
\begin{align}
& \int d^{2}x ~\xi_{A}D_{B}\pi^{BA} = -\int d^{2}x ~\pi^{BA}D_{B}\xi_{A} \notag\\&= -\frac{1}{2}\int d^{2}x ~\sigma_{CD}(\pi^{BC}D_{B}\xi^{D}+\pi^{BD}D_{B}\xi^{C}-\xi^{B}D_B\pi^{CD})-\frac{1}{2}\int d^2x \xi^{B}D_B\pi
\end{align}
In the last equality, the first term can be identified as the lie derivative of the monetum conjugate, while the second term can be identified as the lie derivative of the trace of the momentum conjugate $\pi \equiv \sigma_{CD}\pi^{CD}$.
\begin{align}
\frac{1}{2}\int d^{2}x ~\sigma_{CD}\mathcal{L}_{\xi}\pi^{CD}-\frac{1}{2}\int d^2x \mathcal{L}_{\xi}\pi = -\frac{1}{2}\int d^2x \pi^{CD}\mathcal{L}_{\xi}\sigma_{CD}
\end{align}
Consider now the smeared form of the momentum constraint
\begin{align}
\xi[C_A]:=\int \pi_v\mathcal{L}_{\xi}N_v+\int ~ \pi^{AB}\mathcal{L}_{\xi}\sigma_{AB}\label{smear2}
\end{align}
Computing the poisson bracket of the second term 
\begin{align}
&\{\xi_1[C_A]_{s.t.}, \xi_2[C_A]_{s.t.}\}_{P.B.}=\{\int ~ \pi^{AB}(x)\mathcal{L}_{\xi_1}\sigma_{AB}(x), \int' ~ \pi^{CD}(x')\mathcal{L}'_{\xi_2}\sigma_{CD}(x')\}_{P.B} \notag\\&= \int\int'\big[-\pi^{CD}(x)\mathcal{L}_{\xi_1}\delta(x-x')\mathcal{L}'_{\xi_2}\sigma_{CD}(x')+\pi^{CD}(x')\mathcal{L}_{\xi_1}\sigma_{CD}(x)\mathcal{L}'_{\xi_2}\delta(x-x')\big]
\end{align}
where $s.t.$ stands for 'second term'. The poisson bracket
\begin{align}
&\{\pi^{AB}(x^A, u, v), \sigma_{CD}(x'^A, u, v')\}_{P.B.} = \delta^A_{(D}\delta^B_{C)}\delta^{(2)}(x-x')\delta(v-v')
\end{align}
was used to derive the above. Integrating over the primed variable in the first term and unprimed variable in the second term gives
\begin{align}
&\{\xi_1[C_A]_{s.t.}, \xi_2[C_A]_{s.t.}\}_{P.B.}= \int\big[-\pi^{CD}(x)\mathcal{L}_{\xi_1}\mathcal{L}_{\xi_2}\sigma_{CD}(x)+\pi^{CD}(x)\mathcal{L}_{\xi_2}\mathcal{L}_{\xi_1}\sigma_{CD}(x)\big] \notag\\&= \int \pi^{CD}\mathcal{L}_{[\xi_2, \xi_1]}\sigma_{CD} = -[\xi_1, \xi_2][C_A]
\end{align}
This shows that the momentum constraint forms a closed algebra. If one works in the conformal gauge and make the following choice $\xi_1 = (z^{n+1}, 0) := \xi_n, \xi_2 = (z^{m+1}, 0) := \xi_m$, then
\begin{align}
&\{\xi_n[C_A]_{s.t.}, \xi_m[C_A]_{s.t.}\}_{P.B.}= (n-m)[\xi_{n+m}][C_A]_{s.t.}
\end{align}
since, $[\xi_n, \xi_m] = (m-n)\xi_{n+m}$. The above is precisly the Witt algebra. The same can be shown for the first term of (\ref{smear2}). Hence, it follows that
\begin{align}
&\{\xi_n[C_A], \xi_m[C_A]\}_{P.B.}= (n-m)[\xi_{n+m}][C_A]
\end{align}
\bibliography{QGBib}
\bibliographystyle{SciPost_bibstyle}
\end{appendix}




\nolinenumbers

\end{document}